\documentclass[twocolumn,showpacs,preprintnumbers,amsmath,amssymb]{revtex4}
\usepackage{dcolumn}
\usepackage{bm}
\usepackage{textcomp}
\usepackage{epsfig,bm}

\begin{document}


\title{Vibrational dynamics and boson peak in a supercooled polydisperse liquid}
\author{Sneha Elizabeth Abraham}
\author{Biman Bagchi\footnote[1]{email: bbagchi@sscu.iisc.ernet.in}}
\affiliation{Solid State and Structural Chemistry Unit, Indian Institute of
Science, Bangalore 560 012, India}

\date{\today}

\begin{abstract}

      Vibrational density of states (VDOS) in a supercooled
polydisperse liquid is computed by diagonalizing the Hessian matrix
evaluated at the potential energy minima for systems with
different values of polydispersity. An increase of
polydispersity leads to an increase in relative population of
the localized high-frequency modes. 
At low frequencies, the density of states show an excess compared to the Debye 
squared-frequency law, which has been identified with the boson peak. The height of the boson 
peak increases with polydispersity. 
The values of the participation ratio as well as the level spacing statistics demonstrate 
that the modes comprising the boson peak are largely delocalized. Interestingly, the intensity 
of the boson peak shows a rather narrow sensitivity to changes in temperature and is 
seen to persist even at high temperatures. 
Study of the difference spectrum at two different polydispersity reveals that the increase in the height of boson peak
is due to a population shift from modes with frequencies above the
maximum in the VDOS to that below the maximum, indicating an increase in the fraction of the 
unstable modes in the system. The latter is further supported 
by the facilitation of the observed dynamics by polydispersity.
Since the strength of the liquid increases with
polydispersity, the present result provides an evidence
that the intensity of boson peak correlates positively with the
strength of the liquid, as observed earlier in many experimental systems.
 
\end{abstract}


\maketitle

\section{INTRODUCTION}
	A glass behaves mechanically like a solid but structurally
like a liquid with relaxation times varying from a few minutes to several centuries. The elasticity of a solid is described in terms of phonons, which are quantized vibrational excitations. Propagating acoustic phonon-like excitations have been observed in glasses and glass-forming liquids down to wavelengths comparable to inter-particle distance \cite{sette}. However, their description is rendered difficult 
since glasses lack the translational invariance of crystalline solids. A ubiquitous feature in the physics of glasses is the anomalous behavior of the low-frequency part of the vibration spectrum and the corresponding thermal properties \cite{anomalies}. While the origin of the linear low-temperature specific heat is commonly attributed to the existence of double-well potentials or two-level systems, there is a considerable debate about the so-called {\em boson peak} \cite{nakayama,ShirmacherEPL}. This peak shows up in the vibrational density of states(VDOS), $g(\nu)$ as an excess contribution, compared to the usual Debye behavior[$g(\nu)\propto \nu^{2} $]. 
It is called boson peak because the temperature dependence of its
intensity scales roughly with the Bose-Einstein distribution.
In addition to the presence of boson peak (BP) (or excess density of states) at low frequencies, one also observes a high-frequency exponential tail \cite{chumakov} in the reduced density of states spectrum. 
  
      The interpretation of the boson peak has been a challenge to
experimentalists and theoreticians and is a subject of controversial 
discussions. While some authors attribute it to local/quasi-local vibrations \cite{nakayama, schober, vainer, gurevich}, some others attribute it to collective motions \cite{chumakov, shirmacher}.
A universal mechanism of the
BP formation in glasses was proposed based on the concept of interacting
quasi-local oscillators \cite{gurevich}. Boson peak has also been explained in terms of  
the affine-non-affine crossover at a certain
mesoscopic length scale\cite{barrat}.
It has also been interpreted as the signature of a phase transition in the space of the stationary points of the energy, from a 
minima-dominated phase (phonons) at low energy to a saddle
dominated phase (without phonons)\cite{parisi}.
The boson peak has also been linked
to those motions giving rise to the
two-level-like excitations seen at still lower temperatures \cite{wolynes}. Recently evidence was presented from
numerical studies suggestive of the equality of the boson peak frequency to the Ioffe-Regel limit for transverse phonons above which transverse phonons do not propagate, and the boson peak was attributed to transverse vibrational modes associated with defective soft structures in the disordered state \cite{shintai}.

      Optical heterodyne-detected optical Kerr effect
data on supercooled acetylsalicylic acid and dibutylpthalate display
highly damped oscillations with a period of a few pico seconds
as the temperature is reduced to and below the Mode Coupling Theory
temperature, $T_{MCT}$ \cite{fayer}. The authors interpret this as
the time domain signature of the boson peak and explain that the increased
translational-rotational coupling is responsible for the boson
peak as $T < T_{MCT}$. The main experimental tools used for the
investigation of VDOS and boson peak in glasses
are Raman scattering and the quasi-elastic neutron scattering(QENS).
Recently, single-molecule spectroscopic studies have
also emerged as a potential tool to study the boson peak
in glasses \cite{vainer}. The main difficulty here is the conflicting interpretations regarding the origin of the boson peak. For instance, nuclear inelastic scattering studies of ferrocene probe molecules in toluene, ethylbenzene, dibutylphthalate and glycerol glasses show that
a significant part of the modes constituting the boson peak
is of collective character \cite{chumakov}. However, from the single molecule spectroscopic studies of many single tetra-tert-butylterrylene molecules embedded to amorphous polyisobutylene the authors conclude that the low-energy vibrational excitations have a local character \cite{vainer}. 

      The theoretical as well as experimental studies are done on
systems/samples in the glassy state. However, since the the phonon-like excitations also persist at elevated temperatures in the supercooled liquid regime \cite{sette}, once can study the vibrational dynamics of supercooled liquids via molecular dynamics simulations. Various groups have studied the vibrational dynamics of supercooled liquids \cite{rahman,violette, laird,mazzacurati}, where the focus has been on the characteristics of the high-frequency or low-frequency vibrational modes. The main emphasis of these studies have been on the collective/local nature of the vibrational modes. 

      In this study we investigate the vibrational dynamics and
boson peak in a polydisperse Lennard-Jones liquid. Polydisperse liquid is one of the simplest model systems that exhibit glass transition and can be conveniently studied
via both experiments \cite{WeekSc,glotzerJCP} and computer simulations
as the size distribution of particles prevents crystallization \cite{murarka,sneha,snehapre}. The rest of the paper is organized as follows. In section \ref{comp_details} we describe the model and
computational details. In section \ref{results_dis} we present our results and give detailed discussions on the same. We give our concluding remarks in section \ref{conclusion}.

\section{COMPUTATIONAL DETAILS}\label{comp_details}

	Micro canonical ensemble molecular dynamics (MD) simulations are
carried out in three dimensions on a system of Lennard-Jones (LJ) particles of mean diameter $\overline \sigma$ with polydispersity in both size and mass. The polydispersity in size is introduced by random sampling from the Gaussian distribution of particle diameters $\sigma$,
\begin{equation}
P(\sigma) = \frac{1}{\sqrt{2 \pi}\delta}exp[-\frac{1}{2}(\frac{\sigma - \overline \sigma} {\delta})^{2}]
\end{equation}
The standard deviation $\delta$ of the distribution divided by its mean $ \overline \sigma$ gives a dimensionless parameter, the
polydispersity index, $S$
\begin{equation}
 S = \frac {\delta} {\overline \sigma} 
\end{equation}
The mass $m_{i}$ of particle $i$ is scaled by its diameter,
\begin{equation}
 m_{i} = \overline m(\frac{\sigma_{i}}{\overline \sigma})^3
\end{equation}
We have chosen $\overline m=1.0$. 
The simulations are carried out at different values of the polydispersity index, $S$ but at fixed volume fraction, $\phi = 0.54$.
Three different system sizes were chosen, $N = 256$, $500$ and $864$.
The results are found to be qualitatively the same for the
three different system sizes studied.

The interactions between the particles are given by the
shifted-force LJ potential
\begin{equation}
U_{ij} = 4\epsilon_{ij}[ ( \frac{\sigma_{ij}}{r_{ij}} )^{12} -  ( \frac{\sigma_{ij}}{r_{ij}} )^{6} ]
\end{equation}
where $i$ and $j$ represent any two particles and
\begin{equation}
\sigma_{ij} = ( \frac{\sigma_{i}+\sigma_{j}}{2}) 
\end{equation}
The LJ interaction parameter $\epsilon_{ij}$ is assumed to be 
the same for all particle pairs and set equal to unity.
The particles are enclosed in a cubic box and periodic boundary conditions are used. The cutoff radius $r_{c}$ is chosen to be $2.5 \overline \sigma$.
The time step used for integrating the equations of motion is $0.001$. All quantities in this study are given in reduced units (length in units of $\sigma$, temperature in units of $\frac{\epsilon}{k_{B}}$ and time in units of $\tau=( \frac{\overline m  \overline \sigma^{2}}{\epsilon } ) ^{\frac{1}{2}}$).

The vibrational density of states is obtained from the
Normal Mode Analysis by solving the secular equation,
\begin{equation}\label{secular}
 |{\bf F} -\nu^{2}{\bf I}|=0
\end{equation}
Here $F$ is the mass-weighted potential energy matrix (also
known as the Hessian matrix),
\begin{equation}
 F_{ij}=\frac{V_{ij}}{\sqrt{m_{i}m_{j}}}
\end{equation}
Equation \ref{secular} can be solved to yield a set of
eigen values (the square of the vibrational frequencies, $\nu^{2}$) and corresponding eigen vectors (normal mode displacement vectors), ${\bf e}_{i}$. From the equilibrium liquid configurations generated by the MD simulations, one constructs the potential energy minima or the inherent structure via conjugate gradient minimization. From the potential energy minima, one generates the Hessian matrix, the diagonalization of which would yield the eigen values  and the eigen vectors. The normal modes thus obtained are called the {\em quenched normal modes} (QNM). 

Instantaneous normal modes (INM) can be obtained
by diagonalizing the Hessian constructed for the equilibrium liquid configurations. Since the instantaneous liquid configuration
is not necessarily a potential energy minimum, one gets both
unstable modes (negative eigen values or imaginary frequencies) and 
stable modes (positive eigen values).

\begin{figure}
 \begin{center}
 \epsfig{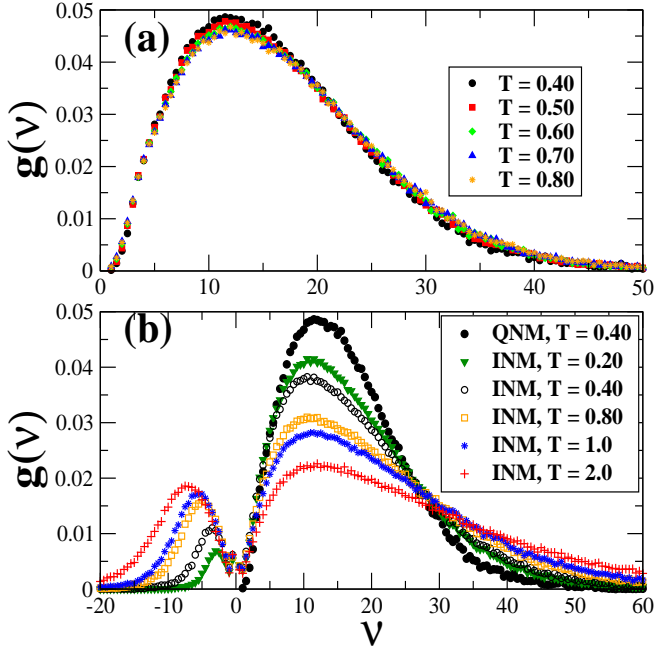}
 \caption{(a)Quenched normal mode spectra at
different values of $T$. The data is for $S=0.20$ and system size $N=256$.
(b)Instantaneous normal mode spectra for different values of $T$. As $T$
decreases the INM spectra approaches closer to the quenched spectrum (dark circles).}
 \label{qnm-inm}
 \end{center}
 \end{figure}

\begin{figure}
 \begin{center}
 \epsfig{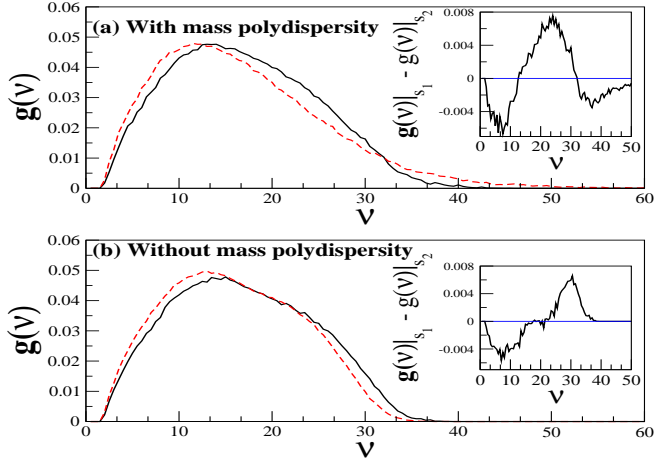}
 \caption{(\textbf{a}) $g(\nu)$ for $S=0.10$ and $S=0.20$ systems and
their difference (inset) (\textbf{b}) Same as (\textbf{a}) but for
$S=0.10$ and $S=0.20$ systems without mass polydispersity. Data shown for a system of $N=256$ particles.}
 \label{diff-dos}
 \end{center}
 \end{figure}

\section{RESULTS AND DISCUSSION}\label{results_dis}

\subsection{Polydispersity Effects on Vibrational Density of States}\label{vdos}

The configuration-averaged vibrational density of states, $g(\nu)$ for
the quenched normal modes is shown in FIGURE \ref{qnm-inm}(a) for
$S=0.20$ at different values of temperature. The width of the
bin chosen to build the histogram is $0.50$.
The VDOS is rather featureless as has been observed by
Rahman et al \cite{rahmanJCP} for single-component Lennard-Jones system.
From FIGURE \ref{qnm-inm}(a), we see that the quenched normal density
of states are only weakly sensitive to the temperature of the parent
liquid. The rather narrow sensitivity of the quenched normal mode spectra to temperature is to be contrasted with the instantaneous normal mode spectra obtained from equilibrium liquid configurations
(See FIGURE \ref{qnm-inm}(b)). The imaginary modes in the INM spectra
are displayed on the negative side of the frequency axis. 
Both the stable (positive frequencies) and unstable (imaginary
frequencies) modes of the INM spectrum show a pronounced change with temperature. As the temperature decreases, the fraction of the unstable modes decreases whereas that of the stable modes increases.  
At low temperature ($T \le 0.40$), the fraction of the
unstable modes is considerably less, indicating that most of the particles are located near the potential energy
minima most of the time. 

Polydispersity has three noticeable effects on the
vibrational density of states.
Firstly, as polydispersity increases, the number of low frequency modes increases whereas the number of high frequency modes
decreases (See FIGURE \ref{diff-dos}). Because of these compensating changes occurring in the high and low frequency regions, one observes
a crossover in the density of states between
$S=0.10$ and $S=0.20$ systems. For the data shown in FIGURE \ref{diff-dos},
this happens at a frequency
$\nu \sim 12.0$ for systems with mass polydispersity and 
$\nu \sim 16.0$ for systems without mass polydispersity.

\begin{figure}
\begin{center}
\epsfig{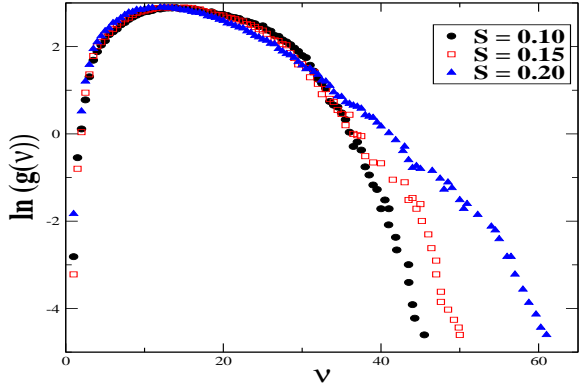}
\caption{Semilog plot of density of states. The data is for $T=0.50$.
The high frequency modes are localized (See Fig \ref{pratio})} 
\label{dos-semilog}
\end{center}
\end{figure}

\begin{figure}
 \begin{center}
 \epsfig{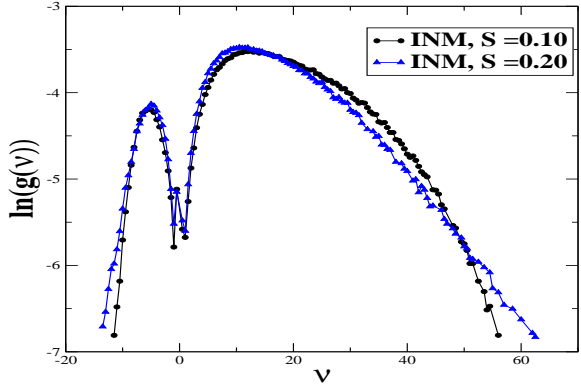}
 \caption{Instantaneous normal mode spectrum for 
$S=0.10$ and $S=0.20$ systems at $T=0.80$.}
 \label{inm-s1s2}
 \end{center}
 \end{figure}

\begin{figure}
\begin{center}
\epsfig{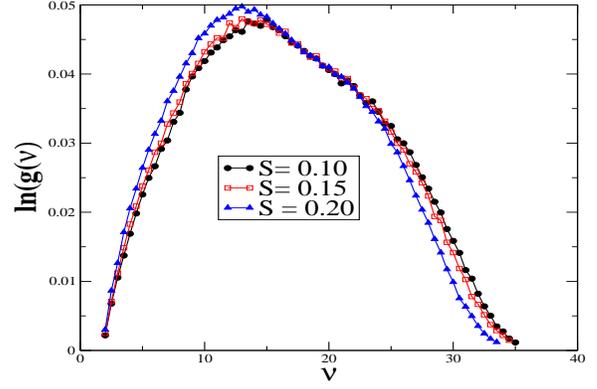}
\caption{Semilog plot of density of states for
different $S$ without mass polydispersity. The data is for $T=0.50$.
}
 \label{dos-semlog-eqmas}
 \end{center}
 \end{figure}

\begin{figure}
 \begin{center}
 \epsfig{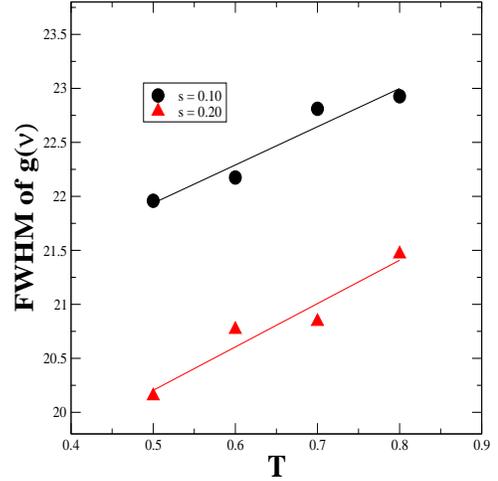}
 \caption{Full width at half maximum, FWHM
for $S=0.10$ and $S=0.20$ as a function of $T$.}
 \label{fwhm}
 \end{center}
 \end{figure}

Secondly, when mass polydispersity is present, there
is a second crossover point in the density of states between
$S=0.10$ and $S=0.20$ systems at a frequency, $\nu \sim 32$ (FIGURE \ref{diff-dos}(a)).
For frequencies higher than this value, there
is an excess of high-frequency modes that increases with $S$. This
is best seen in the semi-log plot of $g(\nu)$ in FIGURE \ref{dos-semilog}.
The plot clearly shows that as polydispersity increases there is
a substantial increase in the number of high frequency modes.
The polydispersity-dependence of the INM spectra is
shown in FIGURE \ref{inm-s1s2}. The $S=0.20$ system has
relatively higher number of high frequency modes
for both the stable and unstable branches.
In FIGURE \ref{dos-semlog-eqmas}, we show the semi-log plot
of VDOS for different $S$ but without mass polydispersity (i.e. all
masses set equal to unity, $m_{i}=1.0$). As we can see from the figure,
the number of high-frequency modes decreases 
with $S$ for systems having no mass polydispersity. 
Thus the excess high frequency modes (whose fraction
increases with increase in polydispersity) is due to the
mass polydispersity effect rather than size polydispersity effect.
(Below we show that the high frequency vibrations are all localized.) 
Thirdly, the vibrational density of states spectrum becomes
narrower with polydispersity.  
In FIGURE \ref{fwhm} we plot the full width at half maximum (FWHM)
of $g(\nu)$ for $S=0.10$ and $S=0.20$ as function of $T$.
The FWHM decreases sharply with $S$ but shows only a weak
increase with temperature. 

\begin{figure}
 \begin{center}
 \epsfig{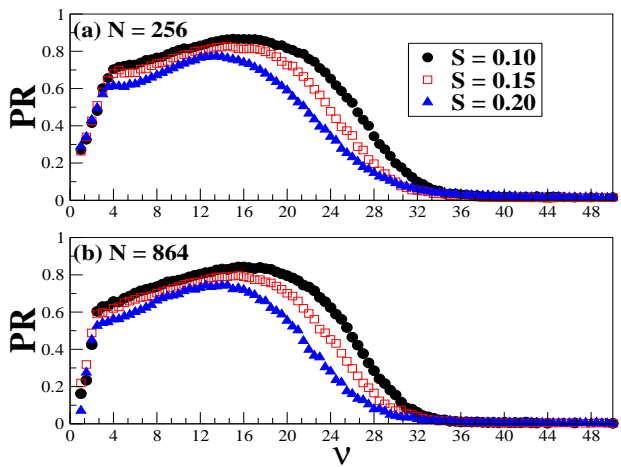}
 \caption{Participation ratio (PR) of the modes for system sizes, (\textbf{a}) $N=256$ and (\textbf{b}) $N=864$. PR gives the
number of particles participating in a given mode. The data is shown for
different $S$ at $T=0.50$. The plot shows that a  $S$ increases, the number
of particles participating in a given mode decreases for all frequencies.
} 
 \label{pratio}
 \end{center}
 \end{figure}

The localization properties of the normal modes can be 
quantified via the participation ratio (PR) which is a measure of the
number of particles participating in a given vibrational mode. 
The participation ratio of mode $i$ is defined as
\begin{equation}
 PR_{i}=[N\sum_{\alpha = 1}^{3N} (e_{i}^{\alpha}.e_{i}^{\alpha})^{2}]^{-1}
\end{equation}
The participation ratio is of the order of $\frac{1}{N}$ for localized modes and is of the order of $1$ for extended modes. 
In FIGURE \ref{pratio}, the participation ratio of the modes is plotted. The averaging is done over modes corresponding to eigenvalues 
in a histogram bin of width $0.50$. As one can see from the
plot, for $\nu > 30.0$, the participation ratio is very low, indicating
that the high-frequency modes are all localized. 
Therefore the high-frequency tail of the normal mode spectrum 
can be attributed to the localized vibrations.
In a crystalline solid, localized vibrations occur due to the
presence of very light impurity atoms or interstitial atoms (that cause
large lattice strains).
It has been shown that when the mass of the impurity atom in a crystal is 
smaller than that of the other particles, the vibrations do not
propagate through the system but gets localized around the 
impurity particle \cite{takizawa}.  

\begin{figure}
 \begin{center}
 \epsfig{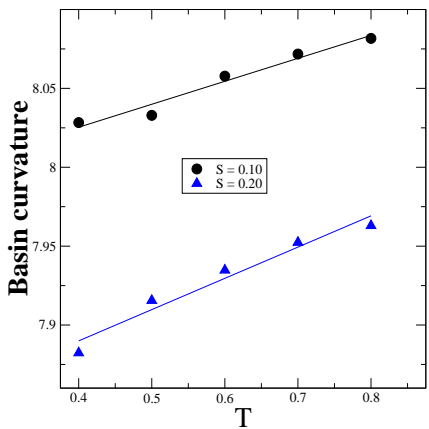}
 \caption{Basin curvature as a function of temperature for $S=0.10$
and $S=0.20$ systems.}
 \label{curvature}
 \end{center}
 \end{figure}

In the frequency range $3.0 \le \nu \le 35 $, one finds
delocalized modes as implied by the high values of PR.
The participation ratio of the modes decreases with polydispersity 
for modes of all frequencies(See FIGURE \ref{pratio}) implying that
the number of particles participating in a mode of given
frequency decreases with $S$. Thus the vibrations
become more localized with $S$. This means that as size disparity among the particles increases, the system cannot sustain propagating modes.
It is interesting to note here that the large size disparity among the particles also leads to the suppression of growth of dynamic 
heterogeneity with polydispersity in supercooled liquids(See \cite{snehapre}). Dynamic heterogeneity in supercooled liquids in its
simplest sense means clusters of fast-moving particles that
move together for a certain amount of time before they get decorrelated.  
When size disparity is large, the particle motion gets decorrelated much faster and hence the formation of \textit{dynamic} clusters is suppressed at higher polydispersity. In the inherent
structure formalism developed by Stillinger and Weber \cite{still-weber,stillinger}, the configuration space of the liquid 
is divided into basins of local potential energy minima(or inherent structures). In the harmonic approximation, which is valid at 
sufficiently low temperatures, each basin is treated 
as a harmonic well. In FIGURE. \ref{curvature} we plot the quantity 
$N^{-1}\sum_{k=1}^{3N-3} log(h\nu_{k})$ as a function of $T$ for
$S=0.10$ and $S=0.20$ systems. This quantity is an indicator
of the average curvature of the basins \cite{mossa}. Since it
is a sum of logarithms it is very sensitive to the tail of the spectrum.
From the plot we see that the average curvature of the basins
increases with $T$ but decreases sharply with polydispersity.
In other words, increasing polydispersity leads to a \textit{flattening}
of the basins which would in turn facilitate inter-basin
transitions and thus enhances the mobility of the system.
The latter has been termed the lubrication effect \cite{sneha}. 

\subsection{Boson Peak}\label{bosonpeak}

\begin{figure}
 \begin{center}
\epsfig{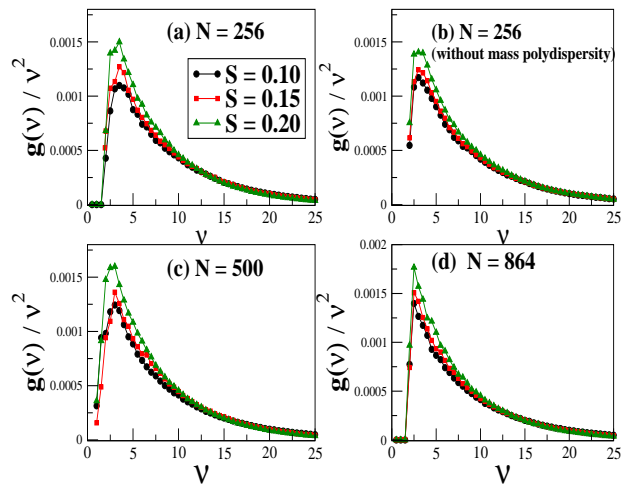}
\caption{Excess density of states (Boson Peak),
$\frac{g(\nu)}{\nu^{2}}$ versus $\nu$. Data shown for different $S$ with the temperature of the parent liquid at $T=0.50$ and for system sizes,
(\textbf{a}) N = 256 (\textbf{b}) N = 256 but without mass polydispersity (\textbf{c}) N = 500 and (\textbf{d}) N=864.}
 \label{bp}
 \end{center}
 \end{figure}

\begin{figure}
\begin{center}
\epsfig{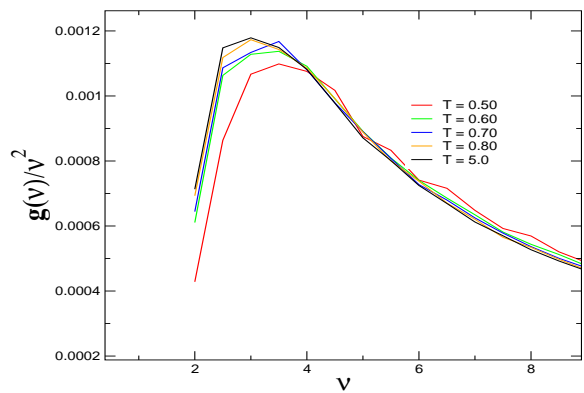}
\caption{Temperature dependence of the Boson Peak.
The data is shown for $S=0.10$ for different temperatures 
of the parent liquid.}
\label{bp-tdep}
\end{center}
\end{figure}

We plot the reduced density of states, $g(\nu)/\nu^{2}$ in FIGURE \ref{bp} for three values of S, namely $S=0.10$, $0.15$ and $0.20$.
All the three systems exhibit the boson peak feature viz. excess density of states as compared to the prediction of the Debye model
($g(\nu)\sim \nu^{2}$). As seen from the figure, the intensity
of the boson peak, $I_{BP}$ increases with $S$ for all the
three system sizes studied. However, \textit{there is no noticeable change in the frequency of the Boson peak, $\nu_{BP}$}.  
The boson peak feature is seen even when we switch off the
mass polydispersity (See FIGURE \ref{bp} (b)) implying that
the size polydispersity effect alone can give rise to the observed
features.
Furthermore, the boson peak feature is seen even for the quenched normal modes obtained from equilibrium
configurations at very high temperature (i.e. $T \gg T_{MCT}$), 
as shown in FIGURE \ref{bp-tdep}. 
As $T$ increases, the 
boson peak height increases but the peak shifts to lower 
frequencies. Such a trend has been observed in experiments
as well \cite{tao,enberg}. However, the temperature-dependence
is weak here. 
The boson peak feature at high temperatures 
shown in FIGURE \ref{bp-tdep} is to be contrasted with the
observation made in an earlier paper\cite{parisi} where the authors
studied the appearance of the boson peak in soft sphere binary 
mixture. In their study, the authors found that as $T$ increases
above $T_{MCT}$, the boson peak feature disappears. 
The authors interpret the boson peak as a manifestation
of an \textit{underlying crossover of the parent liquid's configuration from a saddle dominated dynamics to a minima dominated behavior}.
However, the observation of boson peak at high temperatures makes
such an interpretation questionable. 
\begin{figure}
\begin{center}
\epsfig{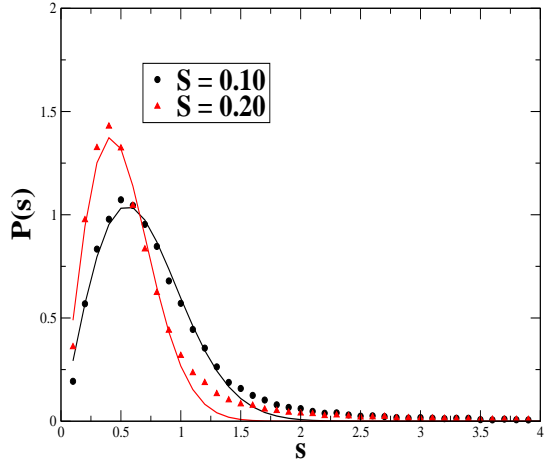}
\caption{The level spacing distribution, $P(s)$ for $S=0.10$
and $S=0.20$ systems for the full eigen value spectrum. Data is shown for $S=0.10$ and $S=0.20$ systems at $T=0.50$. The symbols are the data points and the thick lines are fit to the data according to the
equation, $P(s)=As^{\beta}exp(-Bs^{2})$. The values of $\beta$ are $1.02$ and $1.08$ for $S=0.10$ and $S=0.20$ systems, respectively.}
 \label{lsd}  
 \end{center}
 \end{figure}

The fragility of the current model polydisperse  supercooled liquid
decreases with polydispersity index, $S$ \cite{sneha}. 
This suggests an inverse correlation between the fragility of the liquid and the intensity of the boson peak. 
Such a correlation has actually been observed previously for many liquids in experiments \cite{sokolov} 
as well as in the simulation studies of model glass formers \cite{shintai}.
Shintani and Tanaka \cite{shintai} carried out extensive
simulations on 2D spin liquid glass former in which the
fragility can be varied over an extremely wide range. This was
achieved by varying the anisotropic part of the potential, $\Delta$
which is a measure of the strength of frustration against crystallization. 
As already mentioned, the authors found an inverse correlation between
the fragility and the intensity of the boson peak.
In the current model system the polydispersity index, $S$
plays the same role as $\Delta$. Auer and Frenkel have shown that crystal 
nucleation in a polydisperse colloid is suppressed due to the increase of the surface free 
energy \cite{frenkel} and studies by several groups \cite{pinaki} show that 
the glassy amorphous phase  becomes the equilibrium phase beyond a terminal value of polydispersity.

 If a correlation between the fragility and the intensity of the boson peak exists, then it would mean that
there is also a link between the fast intra-basin vibrational
dynamics and the slow inter-basin diffusive dynamics (See \cite{scopigno}).
However, it appears hard to reconcile such a \textit{link} with the Adam-Gibbs paradigm as the latter predicts 
a relationship between the relaxation time, $\tau$ and 
the configurational entropy, $S_{c}$ ($\tau \sim exp(\frac{A}{TS_{c}})$).
Clearly, on the time scale typical of vibrational dynamics (few pico seconds), the system would not 
have sampled sufficient configurations
as would be necessary to define $S_{c}$. Hence it is difficult to understand the correlation between the intensity of the
boson peak and fragility from this perspective.

It is interesting to note in this connection that Angell et al \cite{angell-jphys} have 
already suggested that
the Boson peak can serve as a signature of configurational excitations
of the ideal glass structure i.e. the topologically diverse defects
of the glassy solid state. This means that the boson peak is 
related to the topographical features of the potential
energy landscape and is thus involved in determining the
fragility of the liquid. As already discussed in Section \ref{vdos}, 
at higher polydispersity we have \textit{flatter} basins (FIGURE \ref{curvature}). 
This is consistent with the decrease in the fragility
with $S$,  as fragile liquids have rugged heterogeneous landscape
whereas strong liquids have smoother landscapes, in accord with
their constant activation energy predicted by their Arrhenius
behavior \cite{debenedetti-stillinger}. 

The strong liquids in Angell's fragile/strong classification are usually network glass formers 
like $SiO_{2}$, $GeO_{2}$ etc., and they appear to lie almost on the opposite spectrum
of our polydisperse liquid system. 
For a polydisperse LJ liquid, the decrease in
fragility with polydispersity is via dynamic facilitation by smaller particles \cite{sneha, snehapre}. 
An obvious manifestation
of the dynamic facilitation is the size-dependent glass transition temperature \cite{sneha}; as temperature is lowered 
the larger particles freeze in first, followed by the smaller ones.
Therefore dynamic facilitation by polydispersity implies that
at higher polydispersity not only the system has smaller barriers to diffusion 
but it also has more relaxation channels available to it.

Here it is interesting to note that Shirmacher at al \cite{shirmacher} have shown that if a
system of coupled harmonic oscillators (with spatially
fluctuating nearest-neighbor force constants on a simple
cubic lattice) is near the borderline of stability a low-frequency
peak appears in $ g(\nu)/\nu^{2}$ as a precursor of the instability.
In their model system, when the amount of the negative force
constants becomes too large, the system becomes unstable and the
boson peak feature shows up. Furthermore, as the fraction of
negative force constants increases the peak intensity
increases and the peak shifts towards lower frequencies.
Instantaneous normal mode analysis shows (see again FIGURE \ref{inm-s1s2}) that the fraction of unstable modes
increases with polydispersity which implies more
pathways for diffusion \cite{violette} at higher polydispersity. Thus the increase in the intensity of boson
peak with polydispersity can be understood in terms of dynamic
facilitation by polydispersity. The relationship with the strength/fragility of the system, however, still 
remains a bit unclear.

One of the key issues in the interpretation of the boson peak feature is whether the modes comprising 
it are localized or extended. 
Figure \ref{pratio} shows that the modes at the frequency range where boson peak appears 
are largely delocalized with high values of participation ratio ($PR > 0.6$). For frequencies less
than $\nu_{BP}$, the participation ratio drops suddenly. This
sudden drop of $PR$ at low frequencies 
has been attributed to finite size effects \cite{mazzacurati}. These
low-frequency modes have been shown to be extended, notwithstanding their low $PR$ values. 
However, note that the frequency of the boson peak
coincides with the frequency below which $PR$ suddenly drops.
The reason for this behavior is not clear to us.
Furthermore, at the boson peak frequency, the $PR$ decreases with $S$
indicative of a correlation between localization of modes
and boson peak intensity. These features are common to all the three system sizes (N=256, 500 and 864)
studied.

\begin{figure}
\begin{center}
\epsfig{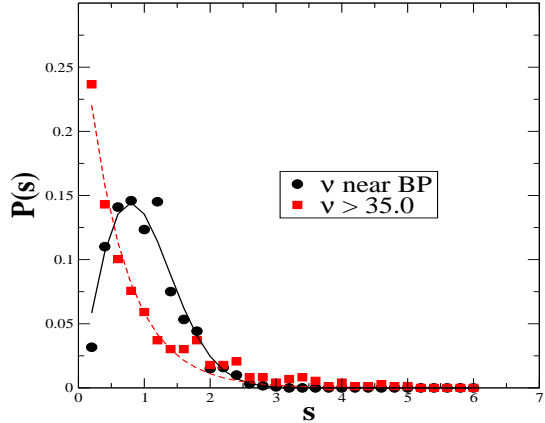}
\caption{The level spacing distribution, $P(s)$ for the eigen values
in the range $3.0 \le \nu \le 4.0$ (near the boson peak) shown in circles and for eigen values in the high frequency range ($\nu > 35.0$) shown by squares. Data is shown for $S=0.10$ at $T=0.50$. The symbols are the data points and the thick lines are fit to the data according to the
the Gaussian orthogonal ensemble (thick line) and
Poissonian distribution (dashed line)}
 \label{lsd-w}  
 \end{center}
 \end{figure}

An alternate way to check whether the boson peak is indeed
associated with localized modes (or not) is by means of level distance 
statistics \cite{shirmacher,izrailev}. The level spacing distribution, $P(s)$ for the 
random-matrix models is defined as the probability of finding the next-nearest-neighbor 
eigenvalue of the spectrum to be at a distance $s$, i.e. $s=\frac{\nu_{i+1}-\nu_{i}}{D} $, where $D$ is 
the mean level spacing. In the case of delocalized states, according
to the Gaussian orthogonal random matrix ensemble, we get
\begin{equation}  
P(s)=As^{\beta}exp(-Bs^{2})
\end{equation}
For localized states, one gets a Poissonian distribution, $P(s)=exp(-s)$.
This is due to the fact that the delocalized states show level repulsion, whereas localized ones do not.
In FIGURE \ref{lsd} we have plotted the level spacing distribution $P(s)$
for the full eigenvalue spectrum along with the fit
for the orthogonal random matrix model.
The values of $\beta$ obtained from the 
fit are $1.02$ and $1.08$ for $S=0.10$ and $S=0.20$ systems, respectively.
This shows that these systems display the behavior according to the Wigner surmise. 
In FIGURE \ref{lsd-w}, we plot the level spacing distribution for frequencies near the 
boson peak ($3.0 \le \nu \le 4.0$) and also for
the high frequencies ($\nu > 35.0$) where the modes are truly
localized (See \ref{pratio}). From the plot it is clear that
statistics at high frequencies follow Poissonian distribution
as expected for localized states. In the vicinity of the boson peak, however, we find a 
Gaussian distribution according to the 
Gaussian orthogonal random matrix ensemble, which means that
the corresponding states are delocalized. However, it should be noted
here that the transverse phonons are shown to be localized
at the boson peak frequency according to the
Ioffe-Regel criterion \cite{shintai, shirmacher}.

\section{CONCLUDING REMARKS}\label{conclusion}

Let us first summarize the main results of this paper.
We have computed the vibrational density of states, $g(\nu)$
in a polydisperse Lennard-Jones system, for three system sizes and three different polydispersity. 
Polydispersity is shown to have  a significant effect on the vibrational density of states. 
Increase in the  polydispersity leads to an increase in the
localized high frequency modes. As polydispersity increases, there is a
softening of vibrational modes manifested in the population shift from modes with 
frequencies above the maximum in $g(\nu)$ to that below the maximum. 
The reduced density of states, $g(\nu)/\nu^{2}$ exhibits the boson peak feature. 
The boson peak is seen to exist even for high
temperature liquid configurations and shows a rather
narrow sensitivity to temperature.
The intensity of the boson peak increases with polydispersity,
which can be understood in terms of dynamic
facilitation by polydispersity by increase in fraction of the unstable modes.

The results clearly show a correlation between an increase in the peak height of the BP with
an increase in the strength of the liquid  -- both are consequences of increase in
polydispersity. This is in agreement with known experimental results.
We find that while the modes comprising the BP are largely delocalized, there is a sharp
drop in the PR of the modes that exist just below the BP. The last observation could be due to
the finite size of the system. The delocalized nature of the BP appears to be quite robust, within
the three system sizes considered here, 
as supported both by the participation ratio and level spacing statistics.  

The observed weak temperature dependence of the boson peak seem to suggest absence of any 
temperature mediated phase transition in the
stationary points of the energy landscape from a saddle dominated to a minima dominated 
regime. However, all the evidences do suggest a close relation between boson peak and unstable modes
of the system. The precise nature of these unstable modes, however, remains unclear at this point.

\begin{acknowledgments}
We thank Professors Chandan Dasgupta and Srikanth Sastry for discussions.
This work was supported in parts by grants from DST, India. 
S. E. Abraham acknowledges CSIR, India for a research fellowship. 
\end{acknowledgments}

\end{document}